# Application-Driven Synthesis and Characterization of Hexagonal Boron Nitride on Metal and Carbon Nanotube Substrates

Victoria Chen[1], Yong Cheol Shin[1], Evgeny Mikheev[2], Joel Martis[3], Ze Zhang[3], Sukti Chatterjee[4], Arun Majumdar[3,5], David Goldhaber-Gordon[2,6], and Eric Pop[1,5,7*]

1. Department of Electrical Engineering, Stanford University, Stanford CA 94305, USA
2. Department of Physics, Stanford University, Stanford CA 94305, USA
3. Department of Mechanical Engineering, Stanford University, Stanford CA 94305, USA
4. Applied Materials, Inc., Santa Clara, CA 95054, United States
5. Stanford Precourt Institute for Energy, Stanford, CA, 94305, USA
6. Stanford Institute for Materials & Energy Sciences, SLAC Natl. Accelerator Lab, Menlo Park, CA, USA
7. Department of Materials Science & Engineering, Stanford University, Stanford, CA 94305, USA

* Contact: epop@stanford.edu

## Abstract

Hexagonal boron nitride (h-BN) is unique among two-dimensional materials, with a large band gap (~6 eV) and high thermal conductivity (>400 $Wm^{-1}K^{-1}$), second only to diamond among electrical insulators. Most electronic studies to date have relied on h-BN exfoliated from bulk crystals; however, for scalable applications the material must be synthesized by methods such as chemical vapor deposition (CVD). Here, we demonstrate single- and few-layer h-BN synthesized by CVD on single crystal platinum and on carbon nanotube (CNT) substrates, also comparing these films with h-BN deposited on the more commonly used polycrystalline Pt and Cu growth substrates. The h-BN film grown on single crystal Pt has a lower surface roughness and is more spatially homogeneous than the film from a polycrystalline Pt foil, and our electrochemical transfer process allows for these expensive foils to be reused with no measurable degradation. In addition, we demonstrate monolayer h-BN as an ultrathin, 3.33 Å barrier protecting $MoS_2$ from damage at high temperatures and discuss other applications that take advantage of the conformal h-BN deposition on various substrates demonstrated in this work.

## 1. Introduction

The family of two-dimensional (2D), layered materials are characterized by relatively strong in-plane bonds and weaker out-of-plane van der Waals coupling between layers. While there has been significant interest in the electrically conductive graphene and semiconducting transition metal dichalcogenides (TMDs), there are also numerous applications for an electrically insulating 2D material.

Hexagonal boron nitride (h-BN) in its monolayer form has a "thickness" of 3.33 Å, taken as the inter-layer spacing of the bulk material [1]. With an electrical band gap of 6 eV, h-BN is an electrical insulator structurally similar to graphene, but composed of ionically bonded boron and nitrogen atoms [2, 3]. This structure, which contains no out-of-plane dangling bonds, results in the h-BN film's high mechanical strength, chemical inertness, and extremely high in-plane thermal conductivity – even greater than the thermal conductivity of bulk copper near room temperature [4-10].

The unique characteristics of h-BN have made it demonstratively useful as the gate dielectric in a 2D field effect transistor (FET) [11], substrate and encapsulant for record high velocity saturation in graphene [12], passivation layer to protect air-sensitive materials [13], and substrate for heat spreading [14]. While mechanical exfoliation from a bulk crystal can yield micron-sized h-BN flakes for proof-of-concept experiments, large area, continuous h-BN films are necessary for practical applications. A promising method to achieve this goal is through low-pressure chemical vapor deposition (LPCVD) at temperatures >900°C [4, 15-17]. Monolayer and multilayer h-BN films have been grown on a variety of substrates, including metals such as copper, platinum, and nickel [16, 18-20]. In addition, h-BN can be deposited on silicon-based substrates ($SiO_2$ and $Si_3N_4$) although the resultant film grain size is limited to ~10-20 μm [21, 22], and by low-temperature electron-enhanced atomic layer deposition (ALD) in nanocrystalline form [23]. These studies have demonstrated that substrate material and crystallinity influence h-BN growth rates, spatial uniformity, and film quality. Researchers have also reported differences in h-BN film properties among different grains of the same growth substrate, another important factor to consider when targeting a uniform, high-quality film for several applications [15, 24].

However, despite growing interest in this material, a direct comparison of monolayer h-BN grown by CVD from an air-stable, solid-source precursor on polycrystalline Pt and single crystal Pt using the same deposition conditions has not been previously reported. Additionally, the characterization of h-BN deposited directly onto both single crystal Pt as well as other unconventional substrates (e.g. carbon nanotubes, CNTs) is relatively limited, even though direct deposition onto CNTs could enable improved device performance as gate dielectrics and CNT protection without requiring the transfer of h-BN. An in-depth study of CVD h-BN on a number of substrates can open doors to a variety of potential applications with unique requirements for the film properties.

Here we report the LPCVD of monolayer h-BN films on single crystal and polycrystalline Pt substrates, as well as multilayer h-BN films on both polycrystalline Cu foil and aligned, single-walled CNTs. From the metal growth substrates, the h-BN films are transferred off and characterized using atomic force microscopy (AFM) and Raman spectroscopy. We additionally show cross-sectional transmission electron microscopy (TEM) images of the crystalline multilayer h-BN films from the Cu substrate and CNTs. Only one previous study [25] has experimentally explored the possibility of CNTs wrapped with few-layer h-BN, whereas here we provide the first demonstration of h-BN capping of CNTs. We also demonstrate, for the first time, the use of monolayer h-BN as an ultrathin capping layer that protects monolayer $MoS_2$ from degradation in high temperature conditions. Finally, we discuss a variety of applications for these ultrathin electrically insulating films.

## 2. Methods

Large-area (on the order of $cm^2$) h-BN films were prepared by LPCVD in a 2" diameter furnace that is schematically represented in figure 1(a). The Pt growth substrates (polycrystalline and single crystal) are placed on a quartz boat inside the furnace chamber and heated to 1100°C, the Cu substrate to 1050°C, and the CNTs on quartz substrate to 1100°C. In each case, the air-stable, solid source precursor, ammonia borane ($H_3NBH_3$), is placed in an ampoule that is heated independently from the main furnace chamber. The ammonia borane is heated to 100°C, at which point it decomposes into borazine [$(HBNH)_3$], polyiminoborane (BHNH), and hydrogen [26]. $H_2$ is used as the carrier gas for the borazine to diffuse through the furnace which is at ~900 mTorr and onto the growth substrates. The metal substrates were annealed at their respective growth temperatures for 40 minutes prior to the h-BN growth, which serves to remove impurities and, in the case of Cu, smooth the substrate surface. The substrates and relative thicknesses of h-BN deposited are schematically summarized in figures 1(b)-(e).

After completing the h-BN growths, we used an electrochemical bubbling method at room temperature to delaminate the h-BN films from the Pt substrates [16], and a wet etching method to transfer them from the Cu substrate to $SiO_2$. A more detailed discussion of the transfer process is included in the supplementary information. This process allows the expensive Pt substrates to be reused for hundreds of growths with no measurable degradation in the substrates or in the grown h-BN quality. Simultaneously, by utilizing this low-temperature transfer process, the h-BN film can ultimately be deposited onto another substrate that is never exposed to the high-temperature growth conditions. Therefore, this process is compatible with applications which contain temperature-sensitive substrates. The majority of samples in this work were transferred onto a 300 nm $SiO_2$ on Si substrate for characterization purposes. Transferring the h-BN films onto the same kind of substrates using the same transfer procedure ensures a fair comparison between films that were originally grown on different substrates.

While the large band gap of h-BN renders it somewhat optically transparent, especially for such thin films, we are able to see some optical contrast on the 300 nm $SiO_2$/Si substrates. The presence of h-BN is additionally verified using Raman spectroscopy, with a 532 nm laser and 100x objective. Bulk h-BN has a Raman peak centered at approximately 1366 cm$^{-1}$. However, thinner h-BN films exhibit blue shifts up to ~4 cm$^{-1}$ with monolayer samples having a peak centered at approximately 1370 cm$^{-1}$ [27]. While this is commonly referred to as the $E_{2g}$ peak, monolayer h-BN belongs to the $D_{3h}$ point group, which differs from the bulk point group ($D_{6h}$). Therefore, we refer to this peak at ~1370 cm$^{-1}$ as the E' peak. In monolayers, the E' peak can shift because of the slightly shorter B-N bonds resulting from the missing interlayer forces that would lengthen B-N bonds in multilayer h-BN [28]. To characterize film thicknesses and compare surface roughness, we use AFM in non-contact mode with a scan rate of 0.5 Hz. We use scanning electron microscopy (SEM) as well as electron backscatter diffraction (EBSD) with an accelerating voltage of 20 kV to examine and analyze the crystallinity of the metal growth substrates. Finally, the samples for cross-sectional TEM imaging were capped with a protective carbon layer and cut with a focused ion beam (FIB).

## 3. Experimental Results and Discussion

The monolayer h-BN growth on Pt substrates is hypothesized to occur by physisorption after the thermal decomposition of the borazine [29]. After the initial monolayer is formed on these surfaces, the surface reactivity decreases and therefore the formation of additional h-BN layers slows, so that the CVD process on Pt is effectively limited to one monolayer at the growth pressure of ~900 mTorr, without the presence of bilayer regions [30]. On the other hand, it should be noted that other works report few-layer regions of h-BN deposited on polycrystalline Pt with different CVD conditions [24, 31]. Therefore, it is evident that the growth mechanisms are highly dependent on variables such as the precursor temperature and pressure of the furnace chamber.

Figure 2 compares the polycrystalline Pt with the single crystal Pt(111) substrate. Figure 2(a) shows a magnified optical image of the polycrystalline Pt foil, with an inset of a photograph of the substrate. Grains on the order of a few hundred μm appear as "sparkles" to the naked eye due to different crystal orientations of the grains reflecting light differently. These grains are also shown by SEM in figure 2(b) and EBSD in figure 2(c). The boundaries between different Pt crystal orientations are very clear in the SEM and EBSD images, and these relatively sharp boundaries indicate that the grain size of the Pt substrate may be a limiting factor in the grain size of the h-BN that is grown. This hypothesis is further discussed in the supplementary information figure S2, which maps the spatial orientation of a monolayer h-BN film that was grown on polycrystalline Pt, revealing clusters of points with the same h-BN orientation in regions of comparable areas to the polycrystalline Pt grain size.

Figure 2(d) shows an optical image of the single crystal Pt(111) substrate, with no grain boundaries visible optically or by the SEM image in figure 2(e). EBSD in figure 2(f) confirms the crystallinity. The surface of the Pt(111) substrate is smoother than the polycrystalline substrate and may demonstrate higher catalytic activity because of its relatively higher surface density of atoms (as compared to other crystal orientations). After characterizing the two Pt substrates, the same conditions (temperature, gas flow, and pressure) are used for CVD of h-BN growth on each. These h-BN films are then transferred with the same method onto 300 nm $SiO_2$ on Si substrates for further characterization.

In Figure 3, we compare monolayer h-BN after transfer from the polycrystalline and Pt(111) substrates to the $SiO_2$/Si substrate. The bubbling-based transfer method has been demonstrated previously in the literature and is summarized in supplementary information figure S1 [16, 32]. Figures 3(a-c) show optical, Raman, and AFM images of the h-BN film transferred from the polycrystalline Pt substrate. Grain boundaries are visible optically in figure 3(a), corresponding to the size and shape of grains from the original metal growth substrate.

Points on different regions of the h-BN film transferred from polycrystalline Pt have different Raman signal intensities as well, which indicate differences in film quality and coverage. This is illustrated in figure 3(b), with points 1A and 1B from the same domain showing very similar signals that both differ from points 2A and 2B in an adjacent domain. The characteristic E' Raman peak of h-BN monolayer can be observed around 1370 cm$^{-1}$ [27]. The additional peak at ~1460 cm$^{-1}$ is due to the Si substrate [22, 33]. By extracting the full width at half maximum (FWHM) values from Lorentzian fits of the Raman data, we also see that the average FWHM from grain 1 is approximately 18.6 cm$^{-1}$, whereas it is 16.9 cm$^{-1}$ in grain 2. These FWHM values have been correlated with the in-plane grain size of the h-BN film, and therefore are evidence of spatial differences in film properties that may arise from different growth substrate grains [34]. Small tears in the film can be seen in the AFM image in figure 3(c) with the underlying $SiO_2$ exposed beneath, and the measured rms surface roughness on the surface of the h-BN film is 1.70 nm.

Figures 3(d-f) show optical, Raman, and AFM images of the h-BN transferred from the single crystal Pt(111) substrate; no grains boundaries are optically visible, but a scratch made in the film provides contrast against the underlying $SiO_2$ substrate. The AFM image in figure 3(e) has a rms surface roughness of 0.80 nm, which is lower than the film from the polycrystalline Pt even though identical transfer methods were used for each film. While surface roughness is an extrinsic measurement that is dependent on AFM scan size, resolution, scan speed, tip sharpness, and other factors, by using

the same conditions for each scan we are able to compare the h-BN films from different growth substrates and conclude that the film from the Pt(111) substrate is smoother.

In the inset shared between figures 3(c) and 3(f), we plot the height distributions of the AFM data for the h-BN films. The h-BN film deposited on the Pt(111) substrate shows a narrower distribution than the one from the polycrystalline Pt substrate, which supports the assertion that the h-BN film is smoother when grown on a single crystal substrate. Raman spectra plotted in figure 3(e) are taken at two different points and have very similar intensities and FWHM values, indicating that the film quality and coverage are much more spatially consistent across the film.

Next, we turn to multilayer h-BN films grown on CNTs/quartz and Cu foil. In figure 4(a), the aligned, single-walled CNTs were grown on a quartz substrate and subsequently placed into the h-BN furnace without transfer [35]. From the TEM cross-section in figure 4(a), it is evident that h-BN selectively deposits onto each individual CNT, but not on the quartz substrate between them. This is also clear in figure 4(b), which shows a cross-sectional TEM of just one CNT that is conformally blanketed with few-layer h-BN. We can refer back to figure 1(e) for a schematic illustration of this geometry. Capping CNTs with h-BN is appealing for improving the electrical performance of CNT transistors; the lack of dangling bonds and surface charge traps make h-BN a promising candidate for higher performance devices (e.g. as gate dielectric), as has been shown with graphene in the past [12, 36]. However, if the h-BN is grown on a different substrate and then transferred to the CNTs, residue left from the transfer process may degrade the transistor performance. Direct growth of h-BN on CNTs is a scalable method of deposition that avoids issues caused by the transfer process.

Finally, in figure 4(c), we show the cross-sectional TEM image of multilayer h-BN that has been transferred to $SiO_2$ from Cu for characterization purposes. These ordered layers in figure 4(c) are uniform across the image, demonstrating the spatial conformity of the deposited h-BN layers. In addition to depositing multilayers as opposed to monolayers with Pt, the h-BN on Cu may be used without transfer in specific applications, as the Cu substrate has a lower cost than Pt and therefore does not need to be reused for growths.

## 4. Applications

Using the films discussed above, we demonstrate the use of monolayer h-BN from the polycrystalline Pt as an encapsulation to protect monolayer $MoS_2$ from anneals at atmospheric pressure under 200 sccm $H_2$ and 200 sccm Ar flow up to 500°C. Figure 5 shows optical images and photoluminescence (PL) spectra for $MoS_2$, as grown and after annealing. The $MoS_2$ is grown on $SiO_2$ on Si using CVD [37], and in figure 5(a) immediately undergoes an anneal at atmospheric pressure in a $H_2$/Ar environment at 500°C; the optical images showing the film before and after clearly illustrate film degradation from the high-temperature environment. Degradation of $MoS_2$ films exposed to elevated temperatures in a hydrogen atmosphere has been previously shown [38, 39] and our PL results confirm this phenomenon. Figure 5(b) shows PL spectra from $MoS_2$ films that are annealed under different conditions, and there is significant quenching of the PL peak even down to anneals at 300°C, indicating damage to the film. On the other hand, by transferring a monolayer of h-BN to blanket the $MoS_2$ prior to the anneal, the $MoS_2$ film is protected from degradation, as shown optically in figure 5(c). In figure 5(d), the PL spectra are retained for anneals up to 450°C, as further evidence that the $MoS_2$ is protected.

In comparison, similar results have been previously achieved with encapsulation by depositing ~15 nm $Al_2O_3$ onto the $MoS_2$ [40]. In other words, the results shown in figure 5 of this work demonstrate an h-BN capping monolayer with similar film protection performance that is only one atomic layer thick – nearly 50 times thinner than the $Al_2O_3$ capping layer previously used. Oxidation prevention methods are a widely researched area, and h-BN is emerging as an extremely thin candidate that can sustain high temperatures while contributing minimal weight to components [41].

To further explore applications of h-BN, we summarize other potential uses in figure 6. As schematically represented in figure 6(a), h-BN may serve as the insulating layer in a metal-insulator-semiconductor (MIS) contact, depinning the Fermi level while acting as a solid-state barrier that prevents metal contacts from reacting with the semiconductor material. This concept has been experimentally demonstrated with both $MoS_2$ and $MoTe_2$ as the 2D semiconductor material [42, 43]. Along these lines, the h-BN lattice structure is highly impermeable to many small chemical species, and therefore is a useful solid-state barrier to prevent undesirable chemical reactions and preserve materials of interest [44]. With excellent measured dielectric properties, it has also been demonstrated as an ultra-thin gate dielectric material for transistors, opening a new potential avenue to explore in the field of transistor scaling [11, 12, 45-47]. This is schematically represented in figure 6(b).

With figure 6(c), we illustrate how multilayer h-BN grown on Cu foil can be used as the switching layer of a resistive random-access memory (RRAM) device, with defects utilized to form the conductive filament [48, 49]. Finally, the high in-plane thermal conductivity of h-BN makes it a promising candidate for heat spreading applications where an electrical insulator is required [14]. Because electrical and thermal conductivity are often positively correlated (e.g. for metals), materials with a high thermal conductivity that are electrical insulators are relatively rare (e.g. just diamond, BN, and AlN) [50]. In addition to this unique property, h-BN also has an anisotropic thermal conductivity between the in-plane vs. cross-plane directions, with a much lower cross-plane thermal conductivity [51]. This would make it valuable in applications where directional heat spreading is important. For example, figure 5(d) shows h-BN as the interlayer dielectric material for a 3D integrated circuit, stacking a memory layer on top of a logic layer. The h-BN can spread heat laterally, and reduce peak hot spot temperatures from logic devices, while blocking the heat from affecting the memory layer above.

## 5. Conclusions

We have demonstrated a scalable method for depositing large-area h-BN films on various substrates (including crystalline Pt, polycrystalline Pt and Cu, and single-wall CNTs) and characterized the resulting films. The crystallinity of the substrates affects the properties of the h-BN as well as the resulting thickness of the film, and this knowledge can be used to selectively tailor the resulting h-BN film properties. In addition, direct growth on CNT and Cu substrates can enable use of the h-BN without a transfer being necessary. We also discuss applications for h-BN grown by CVD and use this material as an ultra-thin barrier layer to effectively protect a monolayer of $MoS_2$, allowing it to reach temperatures above the threshold at which it would typically degrade without sustaining measurable damage. This illustrates promising applications for h-BN as a protective coating against oxidation, which would be of use in numerous industries. While future work remains needed to reduce the growth temperature of h-BN and improve the transfer process when required, this study explores h-BN synthesis and specifies potential target applications that would utilize the carefully tuned properties of the films.

## Acknowledgments

This work was performed at the Stanford Nanofabrication Facility (SNF) and Stanford Nano Shared Facilities (SNSF) supported by the National Science Foundation (NSF) under award ECCS-1542152. This work was supported in part by the Air Force Office of Scientific Research (AFOSR) grant FA9550-14-1-0251, the NSF EFRI 2-DARE grant 1542883, and the Stanford SystemX Alliance. VC acknowledges support from the Stanford Graduate Fellowship (SGF). EP acknowledges partial support from ASCENT, one of six centers in JUMP, a SRC program sponsored by DARPA.

## Notes

The authors declare no competing financial interests.

## Supplementary Information

Additional schematics and details on the procedure to transfer h-BN monolayers from the reusable Pt substrates onto a target substrate. TEM orientation mapping of monolayer h-BN transferred from polycrystalline Pt to a SiN TEM grid.

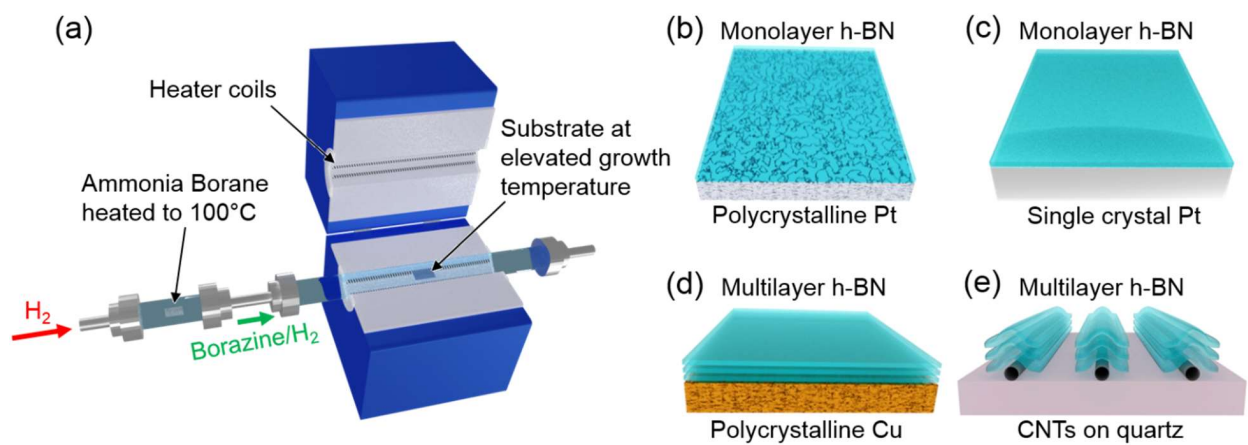

**Figure 1.** Schematics of the CVD furnace and growth substrates. (a) The CVD system used for h-BN growth, (b) monolayer h-BN on a polycrystalline Pt substrate, (c) monolayer h-BN on a single crystal Pt(111) substrate, (d) multilayer h-BN on a Cu substrate, and (e) multilayer h-BN on carbon nanotubes (CNTs) on a quartz substrate.

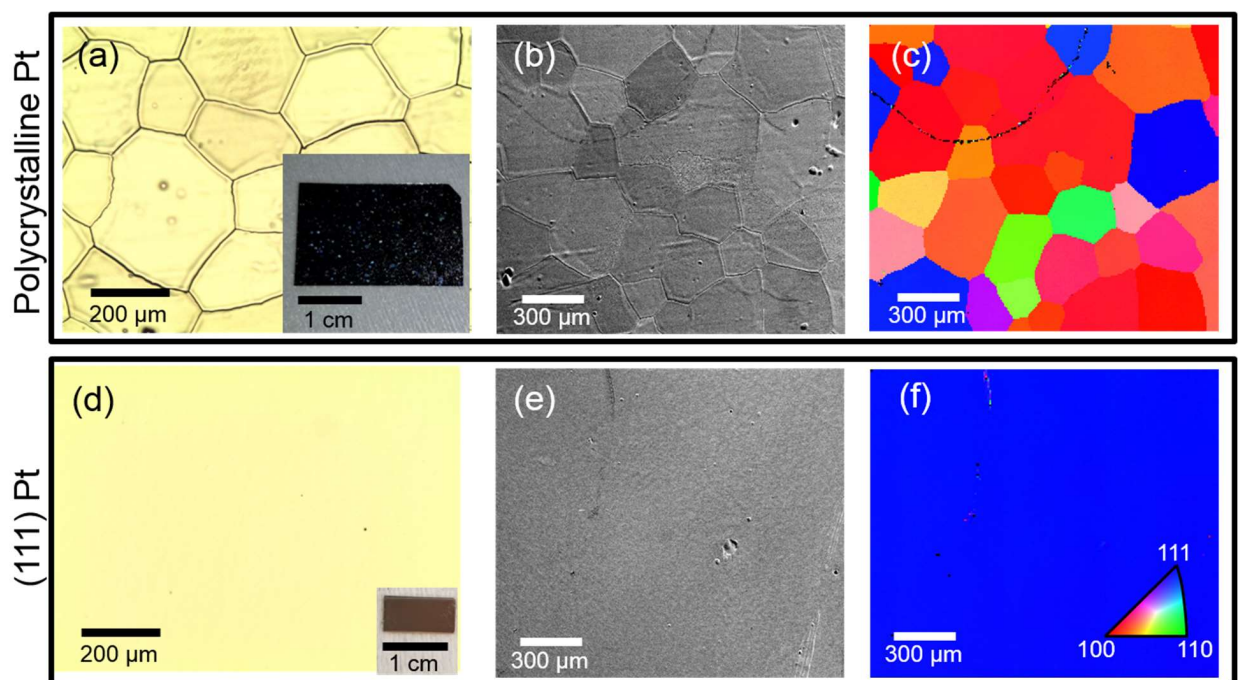

**Figure 2.** Imaging and characterization of Pt growth substrates for h-BN. (a) Zoomed-in optical image of polycrystalline Pt (inset zoomed-out, showing entire foil), (b) scanning electron microscope (SEM) image and (c) EBSD of the same polycrystalline Pt, showing the individual grains and grain boundaries. (d) Zoomed-in optical image of single crystal Pt(111) growth substrate (inset zoomed-out, showing entire sample), (e) SEM image and (f) EBSD of the same single-crystal Pt substrate.

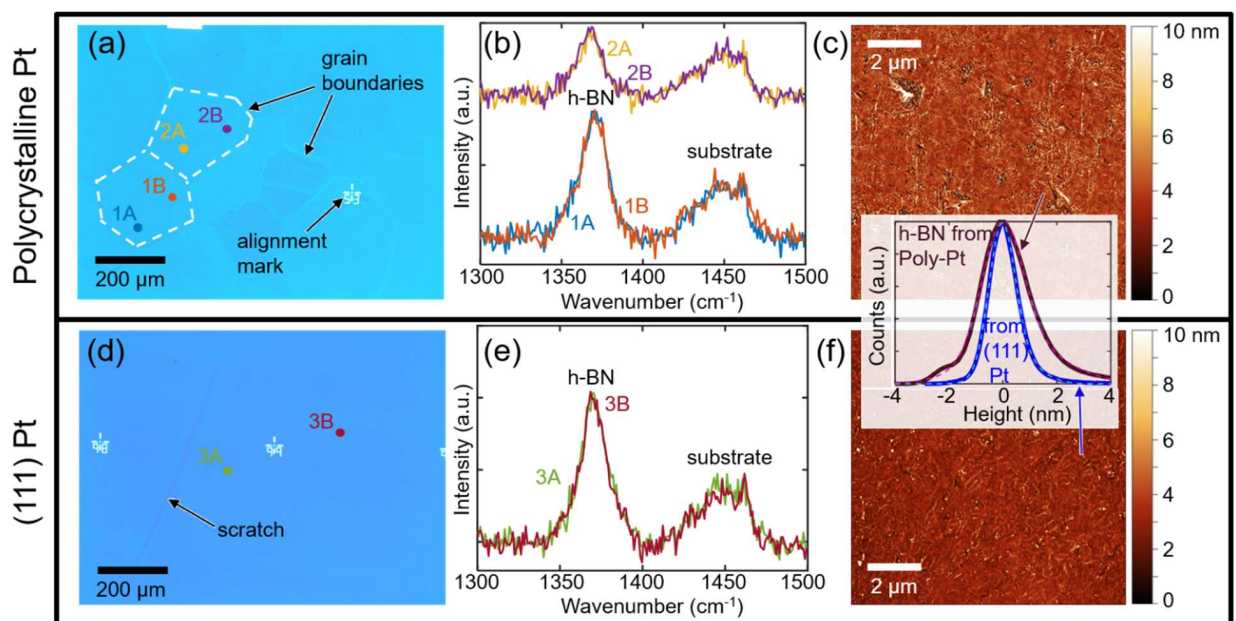

**Figure 3.** Characterization of grown h-BN transferred to $SiO_2$, from growth on (a-c) polycrystalline Pt foil and (d-f) on single-crystal Pt. (a) Optical image, (b) Raman spectra, and (c) AFM topography of h-BN from within a grain marked in (a). Raman signatures are the same within the same grain, but differ between grains. (Spectra offset for clarity.) (d) Optical image, (e) Raman spectra, and (f) AFM topography of h-BN grown on single crystal Pt(111) and transferred to a 300 nm $SiO_2$ substrate on Si. Raman signatures are the same across the uniform h-BN.

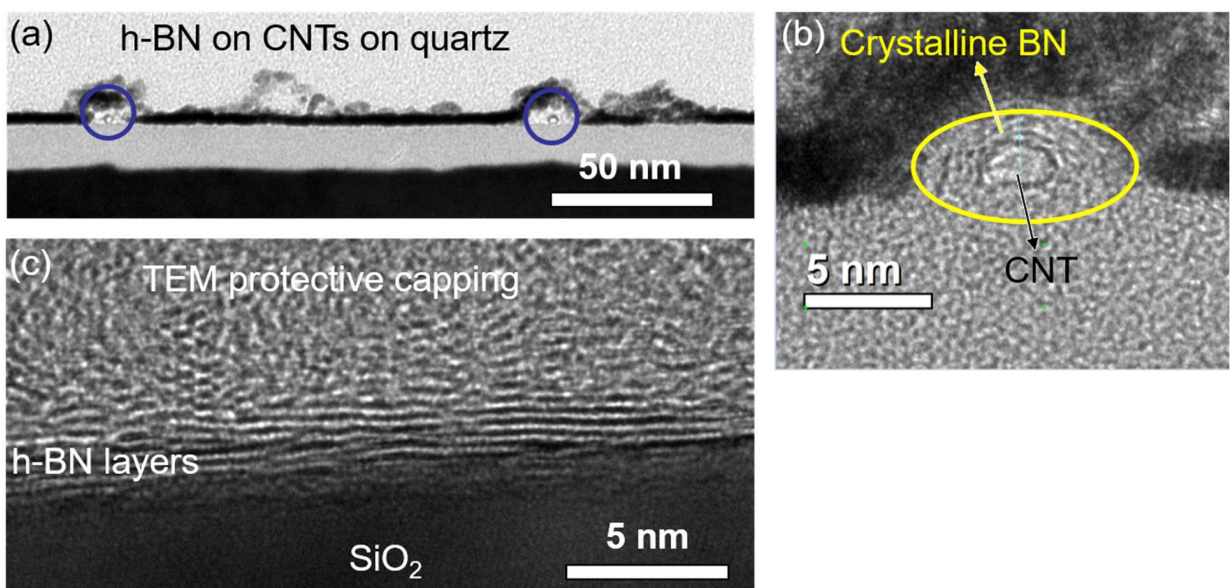

**Figure 4.** Cross sectional TEM images of crystalline, continuous multilayer h-BN grown by CVD on (a) single walled carbon nanotubes, with an image zoomed-in on one h-BN wrapped CNT in (b), and h-BN grown on (c) copper foil (but transferred to $SiO_2$ for imaging).

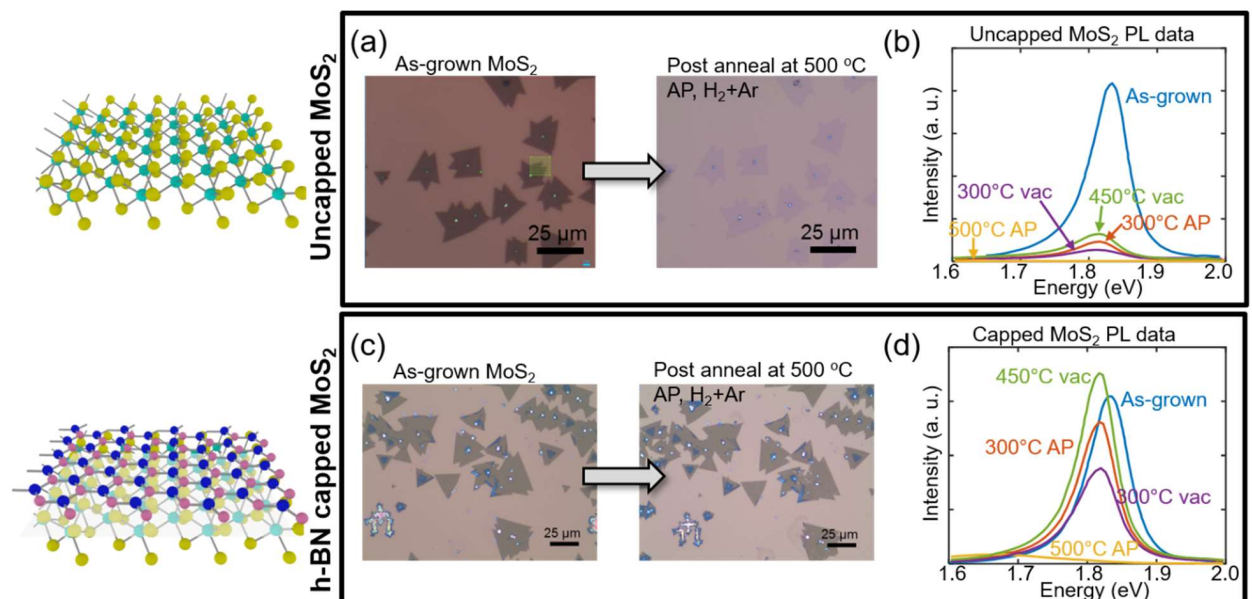

**Figure 5.** Optical and photoluminescence (PL) data on bare vs. h-BN capped $MoS_2$ exposed to elevated temperature conditions. (a) Optical images showing uncapped $MoS_2$ on $SiO_2$ before and after a 500°C anneal at atmospheric pressure (AP) with $H_2$ and Ar, (b) photoluminescence data for as-grown, uncapped $MoS_2$ and $MoS_2$ films after anneals under vacuum (vac) and atmospheric pressure conditions (c) Optical images of h-BN capped $MoS_2$ before and after anneal, and (d) photoluminescence data of the h-BN capped $MoS_2$ after the same anneals.

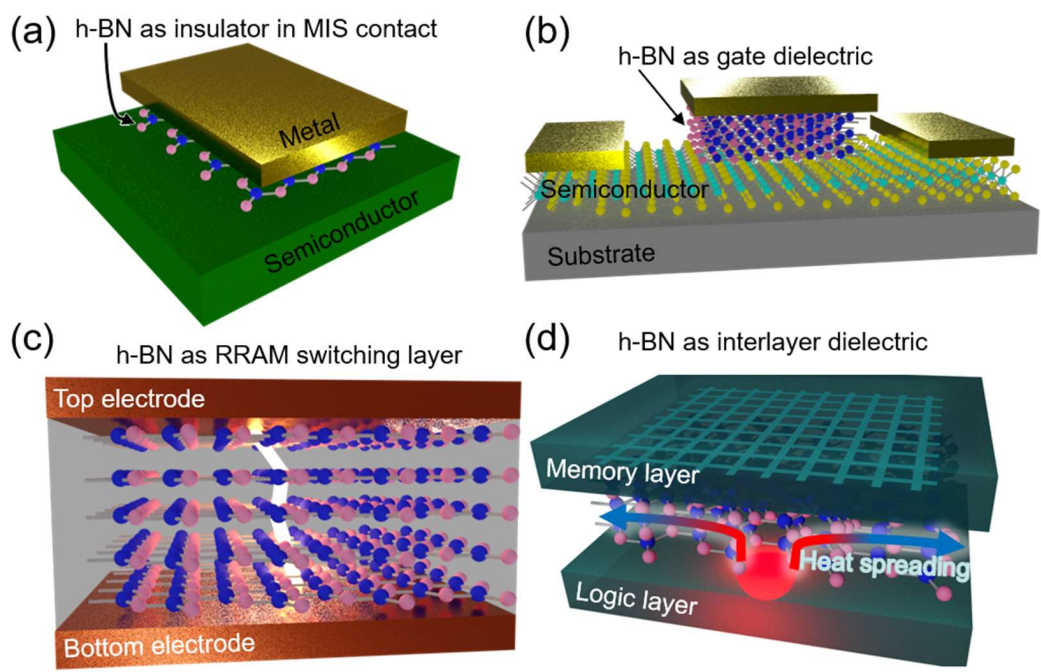

**Figure 6.** Examples of h-BN applications: (a) insulating layer in metal-insulator-semiconductor (MIS) contacts, (b) gate dielectric material, (c) switching layer in RRAM devices, and (d) interlayer dielectric material in a 3D IC to dissipate heat from a hot spot in the logic layer.

SUPPLEMENTARY INFORMATION

# Application-Driven Synthesis and Characterization of Hexagonal Boron Nitride on Metal and Carbon Nanotube Substrates

Victoria Chen[1], Yong Cheol Shin[1], Evgeny Mikheev[2], Joel Martis[3], Ze Zhang[3], Sukti Chatterjee[4], Arun Majumdar[3,5], David Goldhaber-Gordon[2,6], and Eric Pop[1,5,7*]

1. Department of Electrical Engineering, Stanford University, Stanford CA 94305, USA
2. Department of Physics, Stanford University, Stanford CA 94305, USA
3. Department of Mechanical Engineering, Stanford University, Stanford CA 94305, USA
4. Applied Materials, Inc., Santa Clara, CA 95054, United States
5. Stanford Precourt Institute for Energy, Stanford, CA, 94305, USA
6. Stanford Institute for Materials & Energy Sciences, SLAC Natl. Accelerator Lab, Menlo Park, CA, USA
7. Department of Materials Science & Engineering, Stanford University, Stanford, CA, 94305, USA

* Contact: epop@stanford.edu

## A. Transfer Process

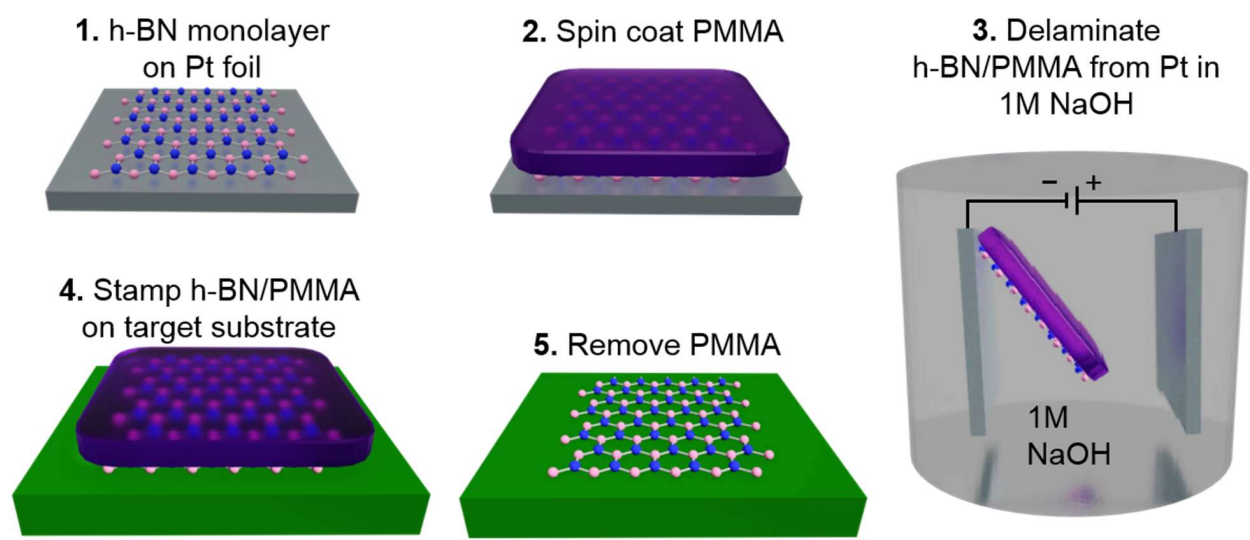

**1.** h-BN monolayer on Pt foil

**2.** Spin coat PMMA

**3.** Delaminate h-BN/PMMA from Pt in 1M NaOH

**4.** Stamp h-BN/PMMA on target substrate

**5.** Remove PMMA

1M NaOH

**Figure S1.** Schematic of transfer process illustrating the delamination of monolayer h-BN deposited onto Pt foil and transferred to the target substrate.

In order to transfer the h-BN monolayer films from the Pt growth substrates for characterization and comparison, we utilize an electrochemical bubbling-based transfer method that has been previously demonstrated [1, 2], as shown in figure S1. After the h-BN is grown on the Pt foil, a 200 nm layer of PMMA is spun onto it and baked at 80°C on a hotplate for 30 min. Then, this stack is subsequently placed in a 1M solution of NaOH and attached to the negative terminal of a power supply, with another Pt foil as the positive electrode. Applying a voltage will generate bubbling at the interface of the h-BN and Pt, allowing the monolayer h-BN capped with the PMMA scaffold to peel off the Pt surface. Once the h-BN/PMMA stack is removed, it is subsequently rinsed in deionized (DI) water and then placed onto the target substrate. Immediately after, a nitrogen spray gun is aimed perpendicular to the sample surface, and the gentle gas flow is used to flatten the film and remove trapped air bubbles. The PMMA is then removed by soaking the sample in acetone for 30 minutes. This relatively clean transfer process allows for the continued reuse of expensive Pt foil substrates, as it does not require any etching and so can preserve the substrate and h-BN films.

## **B. Orientation Mapping**

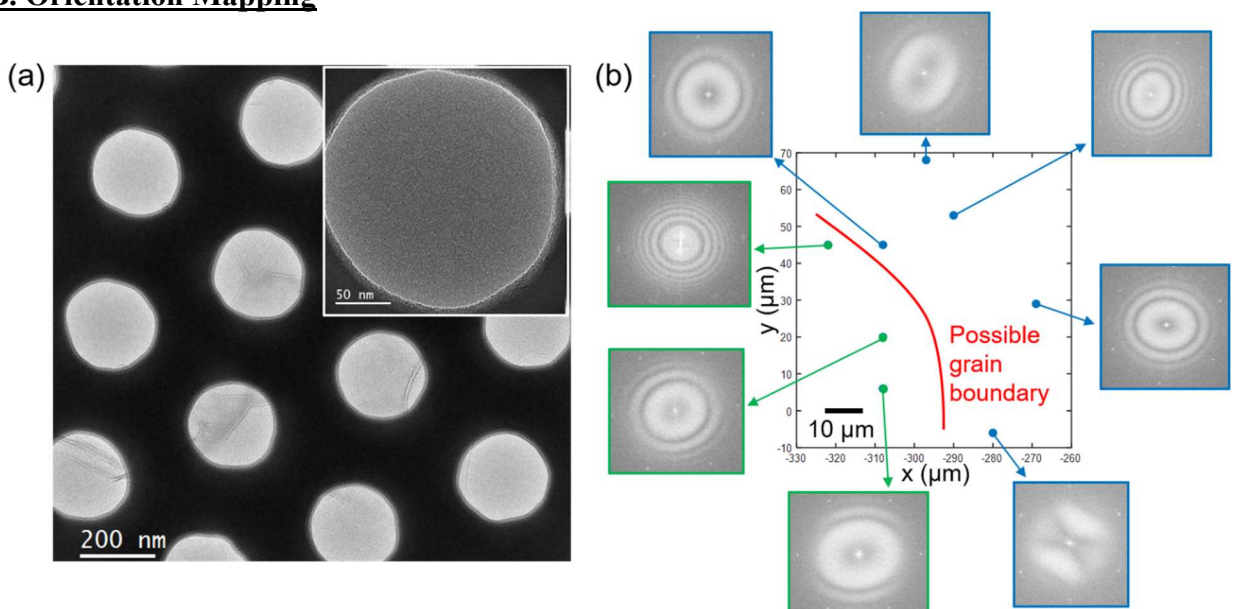

**Figure S2.** TEM images showing diffraction patterns of monolayer h-BN transferred from polycrystalline Pt to a SiN grid. (a) SiN grid with partial coverage of monolayer h-BN (inset showing zoomed-in image of one SiN hole) and (b) spatially mapped diffraction patterns showing similar h-BN orientations in regions of comparable size to the polycrystalline Pt substrate grain sizes.

For TEM imaging, we transfer monolayer h-BN grown on the polycrystalline Pt foil onto a SiN grid, and this grid is pictured in figure S2(a). Then, by examining the diffraction patterns of the h-BN at different points spaced apart from each other in figure S2(b), we observe two distinct clusters of points with two different orientations – one group of points is marked in blue and consistently shows the same orientation in spots up to ~80 μm apart, while the green points mark an adjacent region with a different orientation. A possible grain boundary between the two distinct orientations is sketched with a red line in figure S2(b), and the size of the regions is comparable to the grain size of the polycrystalline Pt growth substrate. Although the number of points imaged was limited due to imperfect h-BN film transfer to these delicate SiN grids, the measured clusters of similar orientations supports the hypothesis that the polycrystalline Pt grain size is a limiting factor in the h-BN domain size.

**Supplementary Information References**